# Understanding Perspectives of Patients, Caregivers and Clinicians towards Emerging Collaborative-decision Making Technologies

**Ray-Yuan Chung, MPH, RD[1], Athena Ortega[1], Zixuan Xu, MS[1], Daeun Yoo, MDE[1], Jaime Snyder, MFA, PhD[1], Wanda Pratt, PhD[1], Aaron Wightman, MD, MA[1], Ryan Hutson, BS[2], Cozumel Pruette, MD, MHS[2], Ari Pollack, MD, MSIM[1]**

**[1] University of Washington, Seattle, WA [2] Johns Hopkins University, Baltimore, MD**

**Abstract**

*In pediatrics, patients, caregivers, and clinicians share responsibility for health decisions, but limited collaboration can undermine outcomes. We conducted a qualitative study examining decision-makers' perceptions toward collaborative decision-making technologies, including interactive dashboards, VR simulators, and AI voice assistants. Findings reveal differences in user opinions across groups and indicate technology acceptance is linked to users' trust of these technologies. Technology developers and researchers need to explore design and implementation strategies that build and facilitate trust or appropriate distrust between users and these novel technologies before these tools can effectively support collaborative decision-making.*

## Introduction

Pediatric chronic care involves complex, ongoing treatment decisions that require active collaboration among youth patients, caregivers, and clinicians. Emerging technologies such as interactive data dashboards, virtual reality (VR) simulators, and artificial intelligence (AI) assistants offer new opportunities to enhance collaborative decision-making. Interactive dashboards integrate patient-reported outcomes with electronic health record data, presenting actionable information through visualizations to support clinician–patient communication and decision-making.[1,2] VR simulators provide a low-risk, immersive, and interactive environment that enhances education and informed decision-making.[3,4] Additionally, recent advances in large language models (LLMs) have strengthened AI assistants, enabling tools such as chatbots and decision-support systems to be integrated into communication among clinicians and families.[5,6] Despite these technological advancements, little is known about how users perceive these technologies or the likelihood that they will be adopted over the long term. To address this gap, we examined the attitudes of patients, caregivers, and clinicians toward these technologies to guide future design and implementation strategies in pediatric chronic care.

## Methods

We conducted seven virtual co-design workshops via Zoom with (1) youth living with chronic kidney disease (CKD), (2) family caregivers, and (3) clinicians. Participants were recruited from a single large children's hospital in the United States. Eligible youth were aged 12–25 years with pre-dialysis CKD stages 3–5, receiving dialysis, or post-transplant; caregivers were eligible if they cared for youth meeting these criteria; and clinicians were eligible if they had experience in pediatric nephrology. Recruitment was conducted through convenience sampling during clinic visits, patient portal outreach, or email. Nineteen participants were enrolled, including six youth (mean age 15.3 years), six caregivers, and seven healthcare providers with an average of 12.1 years of clinical experience (9 years in pediatric nephrology). During the sessions, participants were introduced to three proposed collaborative decision-making technologies through verbal descriptions and AI-generated mockups: (1) a personalized visualization dashboard, (2) a VR simulator for rehearsing clinical conversations, and (3) an AI voice assistant to support in clinic family–clinician interactions. Participants were then invited to share their perspectives and to imagine experiences interacting with the technology. All workshops were audio-recorded, transcribed, and stored on a HIPAA-compliant shared drive. We applied a six-step mixed inductive–deductive thematic analysis approach[7] to examine the transcripts. Two researchers collaboratively developed preliminary codes, which were iteratively refined through multiple rounds of discussion.

## Results

Our findings reveal distinct perceptions toward collaborative-decision making technologies across patients, caregivers and clinicians (Figure 1). Caregivers were the most open to adopting all three technologies, though technology familiarity emerged as a key factor, as illustrated by C3: "*There'll be a learning curve with all of the technologies, especially for me or the older generation.*" While patients identified potential value in each of the presented technologies, some also expressed concerns about the VR and AI tools, "*AI could mix up information from a different type of disease, or somebody else's problems (P6).*" Clinicians also expressed broad concerns about each of the three options, citing redundancy with existing tools and expressing concerns about workflow integration, "*I think the [VR] is so cumbersome, [have to] make it*

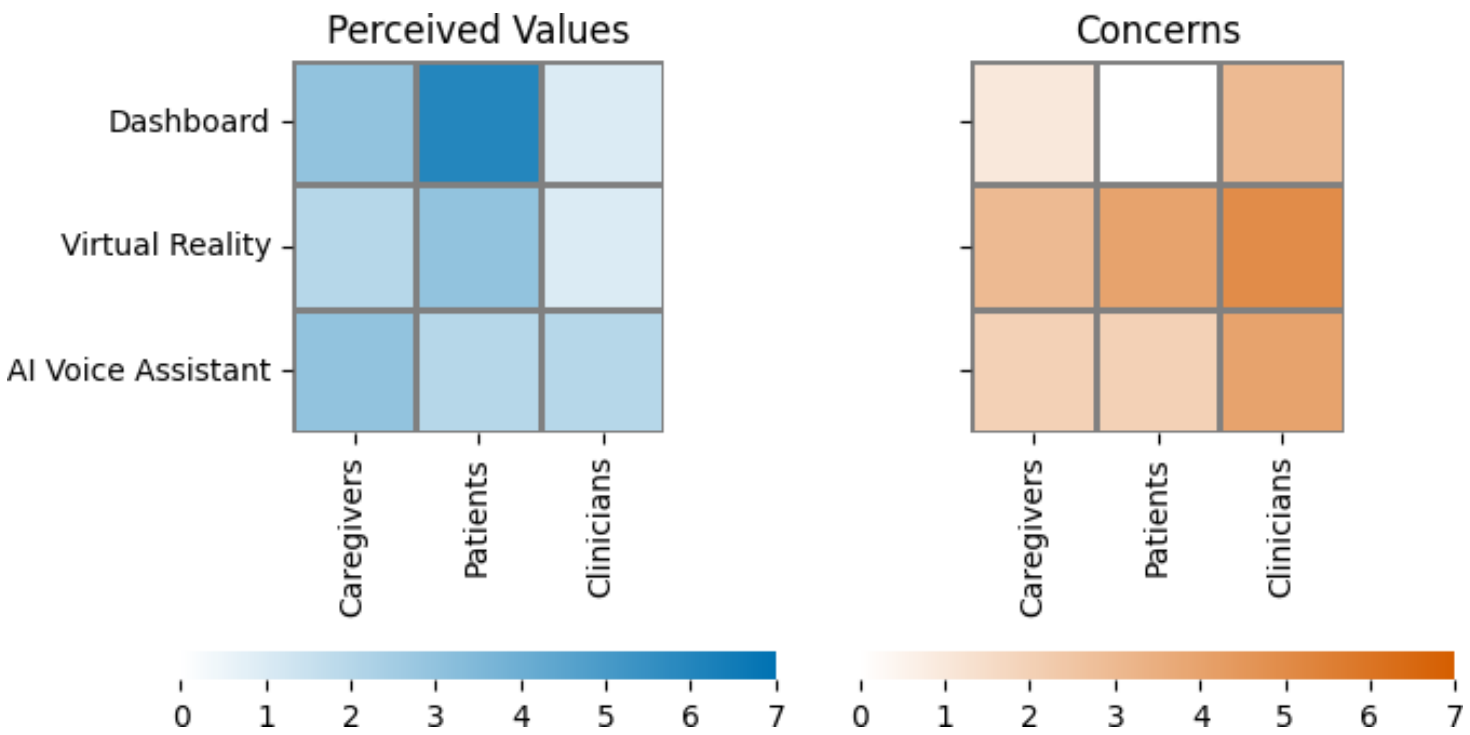

*Figure 1. Heatmaps display the frequency of perceived values (left) and concerns (right) mentioned by each caregivers (C), patients (P), and clinicians (H; i.e., healthcare providers).*

*to the visit and take on and off [at] work (H3)*." Like the patients, clinicians concerns often stemmed from negative attitudes or past experiences with the tools: "*I personally don't like any AI stuff [because] they hallucinate, [and] there could be some bias built into it (H1)*." The themes of perceived values and concerns for all three technologies are listed below:

**Interactive Dashboard**

- *Perceived values*: Improved access to data, better understanding of treatment impacts, data exploration, and enhanced communication between families and clinicians.
- *Concerns*: Potential redundancy with existing technologies, and the learning curve for new health technology platforms could be burdensome.

**VR Simulator**

- *Perceived values*: Safe space to practice communication, facilitate engagement, expressing emotions without real-world consequences.
- *Concerns*: Physical discomfort and fatigue, cost, lack of interaction with real people, cumbersome, and accessibility challenges.

**AI Voice Assistant**

- *Perceived values*: Improved accessibility, support for clinic visit preparation, enhanced communication between families and clinicians, multi-language support, and cost-effectiveness.
- *Concerns*: Distrust due to hallucinations and model bias.

In addition, our analysis revealed that participants' willingness to adopt new technologies appeared to be closely related to their levels of familiarity with these innovations. Dashboard interfaces received the highest levels of enthusiasm, likely because they resemble existing patient portals and visualizations are common in mainstream media, thereby reducing the cognitive burden associated with adoption. In contrast, participants expressed greater uncertainty toward the VR simulator and the AI voice assistant. Uncertainty about how these tools may function may have limited their enthusiasm: "*I'm confused about the AI voice assistant [on] how would that work (C2),*" and "*I wouldn't probably do the VR because it's just a pre-recording (P5)*."

## Conclusion and Future Direction

Our findings reveal one critical factor emerges across all stakeholder groups: trust serves as a fundamental prerequisite for technology adoption in healthcare settings.[8,9] Without established trust, even technologically sophisticated innovations face significant barriers to acceptance and integration into clinical practice. Users' willingness to adopt collaborative decision-making technologies is shaped by their past experiences with similar systems and their confidence in how these tools function.[10] This underscores an important consideration: technological advancement alone may be insufficient, instead trust must be cultivated as it influences adoption, acceptance, and usefulness.[11] While research continues to advance the accuracy and effectiveness of these technologies, addressing trust-related factors including transparency, reliability, and system complexity is essential for successful adoption.[12] Technology designers, developers, and researchers must prioritize building transparent, understandable systems that clearly communicate their functionality and limitations to foster the trust necessary for the collaborative relationships these technologies are designed to support.